%% file: main.tex
\documentclass{article}

\usepackage{microtype}
\usepackage{graphicx}
\usepackage{booktabs} 
\usepackage{hyperref}

\usepackage[accepted]{mlsys2025}

\usepackage{xcolor}
\usepackage{listings}
\usepackage{amsmath}
\usepackage{enumitem}
\usepackage{bookmark}
\usepackage{amssymb}

\definecolor{ghbg}{HTML}{F6F8FA}        
\definecolor{ghkeyword}{HTML}{0366D6}   
\definecolor{ghbuiltin}{HTML}{6F42C1}   
\definecolor{ghstring}{HTML}{22863A}    
\definecolor{ghcomment}{HTML}{6A737D}   
\definecolor{ghnumber}{HTML}{005CC5}    
\definecolor{ghframe}{HTML}{E1E4E8}     

\usepackage{multirow}
\usepackage{siunitx}
\usepackage{threeparttable}
\usepackage{tabularx}
\usepackage{subcaption}

\usepackage{soul} 
\usepackage{adjustbox} 
\definecolor{myyellow}{HTML}{F4AE01}
\definecolor{myblue}{HTML}{0064D3}
\definecolor{mygreen}{HTML}{88B719}
\definecolor{myred}{HTML}{E53238}

\definecolor{darkgreen}{rgb}{0.0, 0.6, 0.0}
\definecolor{darkred}{rgb}{0.7, 0.0, 0.0}

\usepackage{capt-of}   

\newcommand{\cmark}{\textcolor{darkgreen}{\checkmark}}
\newcommand{\xmark}{\textcolor{darkred}{\texttimes}}

\mlsystitlerunning{Meeting SLOs, Slashing Hours: Automated Enterprise LLM Optimization with OptiKIT}

\begin{document}

\twocolumn[
\mlsystitle{\raisebox{-0.55\height}[0pt][0pt]{\includegraphics[height=3.0em]{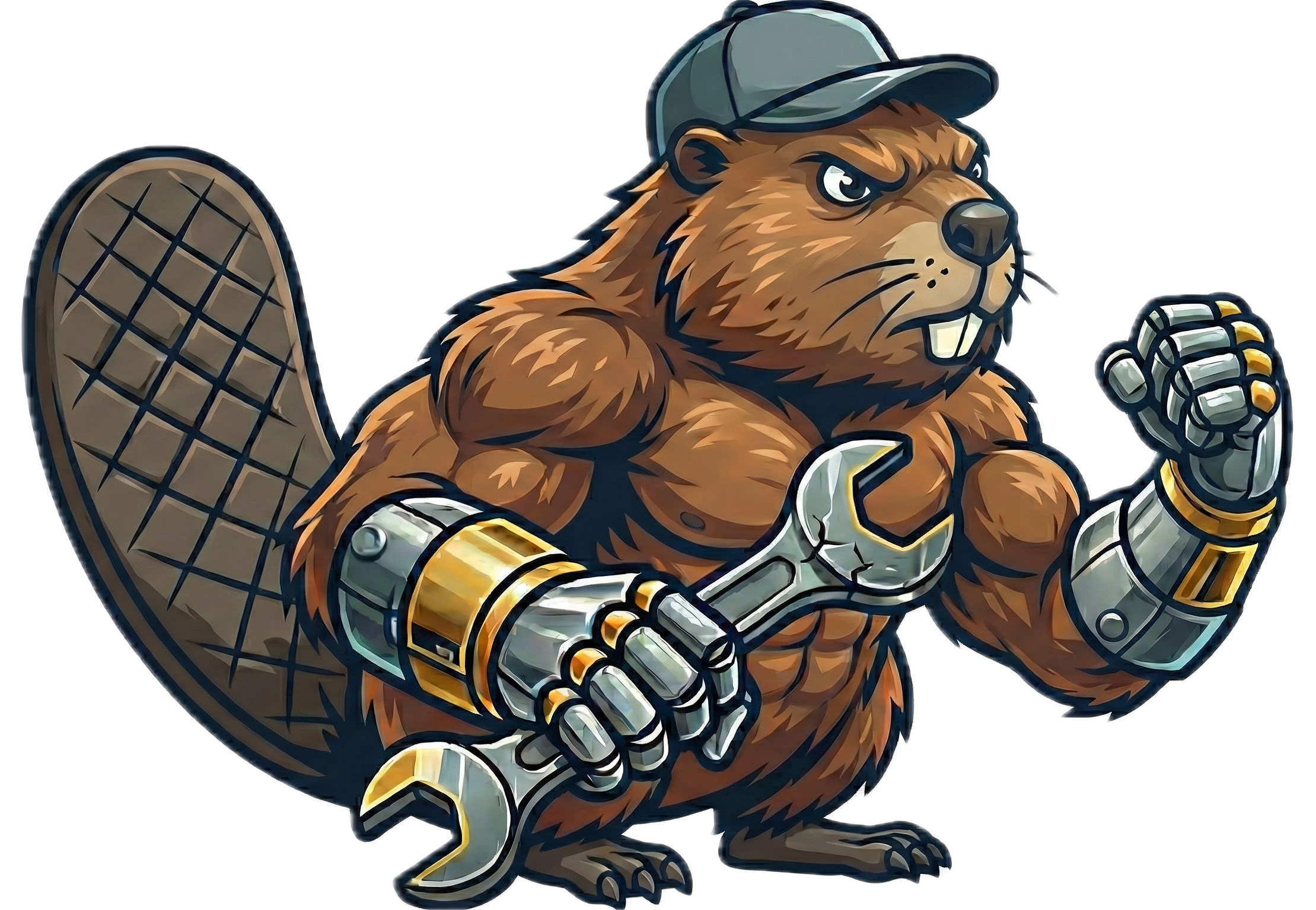}} Meeting SLOs, Slashing Hours: Automated Enterprise LLM Optimization with OptiKIT}



\mlsyssetsymbol{equal}{*}


\begin{mlsysauthorlist}
\mlsysauthor{Nicholas Santavas}{ebay,duth}
\mlsysauthor{Kareem Eissa}{ebay}
\mlsysauthor{Patrycja Cieplicka}{ebay}
\mlsysauthor{Piotr Florek}{ebay}
\mlsysauthor{Matteo Nulli}{ebay}
\mlsysauthor{Stefan Vasilev}{ebay}
\mlsysauthor{Seyyed Hadi Hashemi}{ebay}
\mlsysauthor{Antonios Gasteratos}{duth}
\mlsysauthor{Shahram Khadivi}{ebay}

\end{mlsysauthorlist}

\mlsysaffiliation{ebay}{eBay, Foundation Models Team, Amsterdam, Netherlands}
\mlsysaffiliation{duth}{Democritus University of Thrace, Xanthi, Greece}

\mlsyscorrespondingauthor{Nicholas Santavas}{nsantavas@ebay.com}


\mlsyskeywords{Machine Learning, Inference Optimization}

\vskip 0.3in

\begin{abstract}
Enterprise LLM deployment faces a critical scalability challenge: organizations must optimize models systematically to scale AI initiatives within constrained compute budgets, yet the specialized expertise required for manual optimization remains a niche and scarce skillset.
This challenge is particularly evident in managing GPU utilization across heterogeneous infrastructure while enabling teams with diverse workloads and limited LLM optimization experience to deploy models efficiently.
We present \textsc{OptiKIT}, a distributed LLM optimization framework that democratizes model compression and tuning by automating complex optimization workflows for non-expert teams. \textsc{OptiKIT} provides dynamic resource allocation, staged pipeline execution with automatic cleanup, and seamless enterprise integration.
In production, 
it delivers more than 2× GPU throughput improvement while empowering application teams to achieve consistent performance improvements without deep LLM optimization expertise. We share both the platform design and key engineering insights into resource management, pipeline orchestration, and integration patterns that enable large-scale, production-grade democratization of model optimization. Finally, we open-source the system to enable external contributions and broader reproducibility.
\end{abstract}
]



\printAffiliationsAndNotice{}  

\input{sections/01_introduction}

\input{sections/02_background}

\input{sections/03_system_design}

\input{sections/04_experiments}
\input{sections/05_conclusion}

\bibliographystyle{mlsys2025}
\bibliography{optikit_references}

\newpage
\appendix
\input{sections/0X_appendix}

\end{document}

%% file: sections/01_introduction.tex
\section{Introduction}
\label{sec:introduction}

The proliferation of Large Language Models (LLMs) \citep{brown2020languagemodelsfewshotlearners, grattafiori2024llama3herdmodels, yang2025qwen3technicalreport} across enterprises has created a major computational challenge \cite{chavan2024faster}. As organizations adopt generative AI, they face a fundamental tension between the exponential growth in demand for AI-driven features and the finite and expensive supply of specialized GPU infrastructure. This scalability issue, if unaddressed, threatens to stifle innovation and render the widespread deployment of powerful LLMs economically untenable. At global technology companies, like eBay, this is not a distant prospect but an immediate operational reality. The ambition to enhance user experience with a new generation of LLM-powered applications is constrained by hardware capacity and operational efficiency. Deploying models from 8B to over 70B parameters creates a significant strain on computational resources. Manual optimization \cite{zhu2024survey}, a specialized craft practiced by few experts, does not scale, while existing tools often lack the robustness and seamless integration required for production systems \cite{park2025survey}. This gap forces a trade-off between feature velocity and performance, creating dependencies on a small pool of experts.

\begin{figure}[t!]
\centering
\includegraphics[width=0.45\textwidth]
{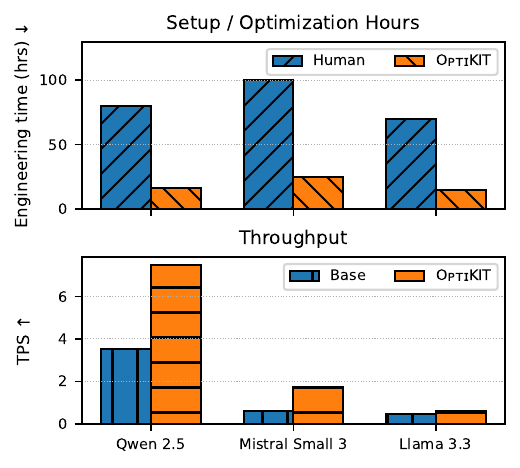}
\vspace{-10pt}
\caption{\textbf{\textsc{OptiKIT} time and throughput gains.} The \emph{top figure} shows the engineering time saved in model optimization through \textsc{OptiKIT} vs internally estimated human hours. In the \emph{bottom figure}, the optimal TPS (Transactions Per Second, i.e., Throughput) after the \textsc{OptiKIT} cycle has terminated vs the baseline. Results shown on three models.}
\label{fig:page1-figure}
\end{figure}


\begin{figure*}[h]
\begin{center}
\includegraphics[width=0.85\textwidth]{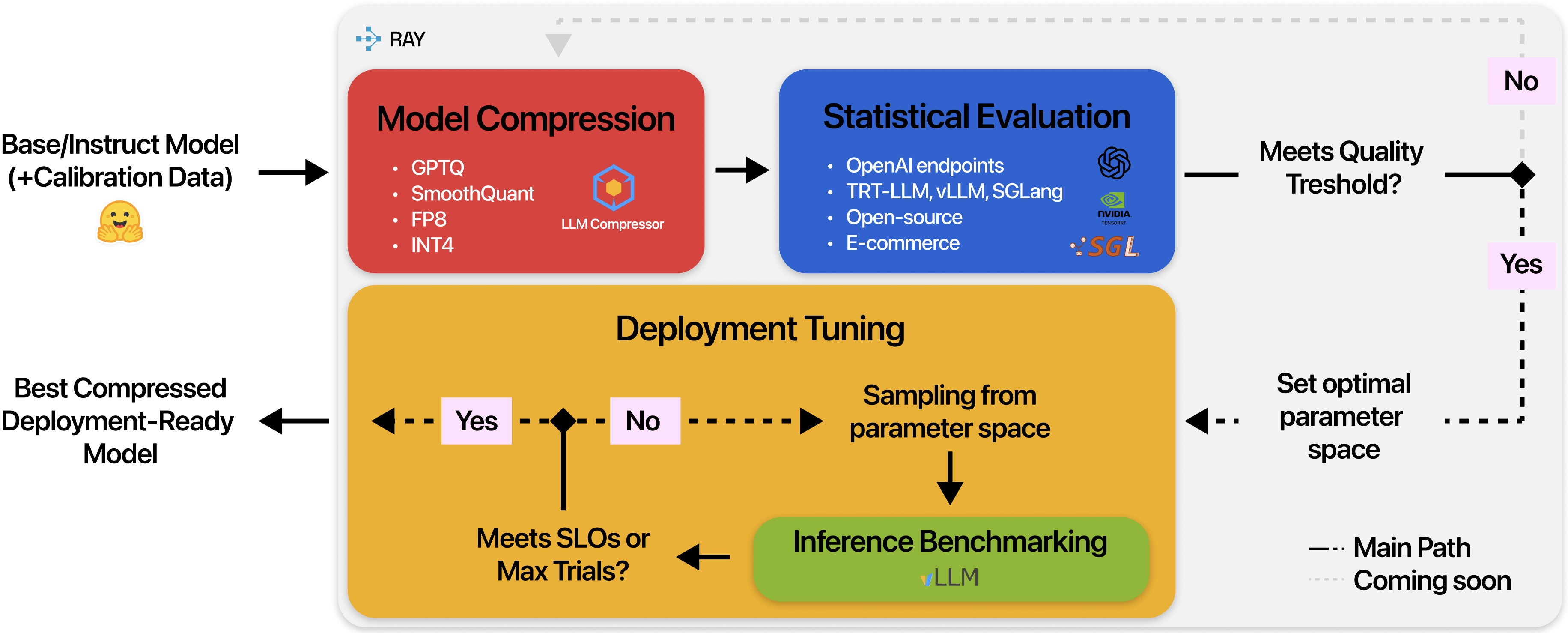}
\end{center}
\vspace{-10pt}
\caption{\textbf{\textsc{OptiKIT} full pipeline.} The figure shows the full \textsc{OptiKIT} flow. We begin by fetching any base/instruct model along with calibration data if needed, and apply model compression through the user-selected technique. We then proceed to perform a statistical evaluation of the optimized model to ensure the validity of our compression strategy. If the performance is up to standards, we determine the set of parameter space for deployment tuning. Subsequently, we sample from this space and perform Inference Benchmarking to determine the optimal subset of parameters for deployment. If the SLOs (Service Level Objectives) are not met, we iteratively repeat the above, sampling a new set of parameters. When a parameter configuration meets the SLOs, we return the model configuration along with its weights, ready for deployment. The small logos represent part of the back-ends supported.}
\label{fig:main_fig}
\end{figure*}

In this paper, we introduce \textsc{OptiKIT}, an automated LLM optimization framework designed to address these challenges. Developed and deployed at eBay,
\textsc{OptiKIT} embodies three core principles: automation, standardization, and deep enterprise integration. It provides a comprehensive, end-to-end solution that automates the optimization lifecycle, from model analysis and resource allocation to performance benchmarking and deployment. By standardizing this process, \textsc{OptiKIT} democratizes access to advanced optimization techniques, enabling any engineering team to achieve expert-level performance without requiring specialized knowledge.

Initial results at eBay demonstrate that \textsc{OptiKIT} can achieve significant throughput gains and latency reductions, enabling the deployment of more powerful models within existing resource envelopes (Figure \ref{fig:page1-figure}). This suggests that a systematic, automated approach to LLM optimization is a technically feasible and critical component for enabling scalable, cost-effective AI 
in the enterprise \cite{zhen2025taming}.

Our contributions can be summarized as follows: 
\begin{itemize}[noitemsep, topsep=0pt]
    \item We present \textbf{\textsc{OptiKIT}}, a fully \textit{automated end-to-end}, distributed, resource-aware LLM optimization pipeline with modular orchestration, dynamic GPU allocation, completely integrated into enterprise infrastructure.
    \item We \textbf{operationalize} this architecture through tailored algorithmic decision-making governed by strict Service Level Objectives (SLOs). Key mechanisms include: a backend-agnostic compression engine paired with a statistical evaluation mechanism acting as a quality-feasibility gate; regression-based stability detection for rigorous throughput certification under explicit serving profiles; and a Bayesian runtime tuner that performs sample-efficient search over costly trials, backed by complete artifact archival for auditability and replay.
    \item We conduct an \textbf{extensive empirical study} across \textit{large-scale production workloads and model families}. \textsc{OptiKIT} achieves throughput gains of up to \textit{$2.8\times$} with robust, reproducible optimization across heterogeneous infrastructure.
\end{itemize}

%% file: sections/02_background.tex
\begin{table*}[htbp]
\centering
\small
\caption{\textbf{\textsc{OptiKIT} vs similar techniques.} We compare \textsc{OptiKIT} with similar LLM optimization techniques.}
\vspace{-8pt}
\begin{adjustbox}{max width=\textwidth}
\begin{tabular}{lcccccc}
\toprule
\textbf{Technique} & \textbf{Model} & \textbf{Quality} & \textbf{Inference} & \textbf{Deployment} & \textbf{Data} & \textbf{Production} \\
& \textbf{Compression} & \textbf{Gate} & \textbf{Benchmarking} & \textbf{Tuning} & \textbf{Sovereignty} & \textbf{Ready} \\
\midrule
Manual optimization \cite{zhu2024survey} & \cmark & \xmark & \xmark & \xmark & \cmark & \xmark \\
TensorRT-Sweep \cite{tensorrtsweep} & \xmark & \xmark & \cmark & \cmark & \xmark & \cmark \\
GuideLLM \cite{guidellm2024} & \xmark & \xmark & \cmark & \xmark & \cmark & \xmark \\
High-Throughput LLM Inference \cite{xiong2025highthroughputllminferenceheterogeneous} & \xmark & \xmark & \cmark & \cmark & \xmark & \xmark\\
SCOOT \cite{cheng2025scootsloorientedperformancetuning} & \xmark & \xmark & \cmark & \cmark & \xmark & \cmark\\
\textbf{\textsc{OptiKIT}} (ours) & \cmark & \cmark & \cmark & \cmark & \cmark & \cmark \\
\bottomrule
\end{tabular}
\end{adjustbox}
\label{tab: existing-tools-comparison}
\vspace{-5pt}
\end{table*}

\section{The LLM Optimization Challenge at Scale}
\label{sec:background}

The deployment of Large Language Models in production environments presents a unique set of challenges that differ substantially from academic 
research settings \cite{chavan2024faster}. Enterprise deployments must contend with hard constraints on GPU availability, heterogeneous hardware infrastructure, and 
the need for consistent performance across diverse workloads. 
At eBay, these challenges manifest in several key areas.
First, the finite nature 
of GPU resources creates a zero-sum constraint where every inefficiency in one application directly impacts the capacity available for others. 
Hence, resource utilization efficiency is paramount. Second, the diversity of model architectures and use cases—ranging from 8B to over 70B parameters—demands flexible optimization approaches.
Manual LLM optimization \cite{zhu2024survey} represents a significant organizational bottleneck. The process requires deep expertise in model compression, hardware-specific 
optimizations, and inference runtime tuning \cite{zhou2024survey2} —knowledge that is concentrated among a small number of specialists. This expertise gap creates several 
problems: optimization work becomes a dependency that slows feature development; inconsistent approaches lead to suboptimal resource utilization; 
and the manual nature of the process introduces variability in outcomes.

We separate engine-specific tooling from the end-to-end deployment workflow. Engine-specific tools operate inside a particular serving stack (profiling, benchmarking, and parameter sweeps), while the workflow coordinates the steps that turn a base model into a deployment-ready, SLO-validated artifact (compression, quality checks, benchmarking, tuning, and archival). Table~\ref{tab: existing-tools-comparison} compares native workflow capabilities; a \cmark indicates the tool provides that function itself rather than being wrapped by external orchestration (\xmark).

Existing tools for LLM optimization, while powerful, often fall short in enterprise settings; see Table~\ref{tab: existing-tools-comparison}.
A key distinction is 
\textit{engine-layer} tooling (profilers, kernel/engine autotuners, and parameter sweeps) versus \textit{workflow-layer} orchestration.
Engine-layer tools (e.g., TensorRT-Sweep \cite{tensorrtsweep}) can efficiently explore inference configurations, but typically do not provide end-to-end quality gating, SLO certification under an explicit serving profile, or complete archival of artifacts/metrics for auditability.
Similarly, benchmarking frameworks (e.g., GuideLLM \cite{guidellm2024}) provide valuable steady-state measurements, but they are not a substitute for a production workflow that (i) filters candidate models via statistical quality checks, (ii) searches for the maximum sustainable rate that satisfies deployment SLOs, and (iii) packages the resulting model, configuration, and evidence for rollout. \textsc{OptiKIT} is designed to orchestrate this workflow; in this paper we use our own benchmarking stack (vLLM-based) and do not integrate external engine-layer tools.

These challenges underscore the need for a unified, production-grade optimization framework— a role \textsc{OptiKIT} aims to fulfill.


%% file: sections/03_system_design.tex
\begin{figure*}[ht]
\centering
\includegraphics[width=0.95\textwidth]{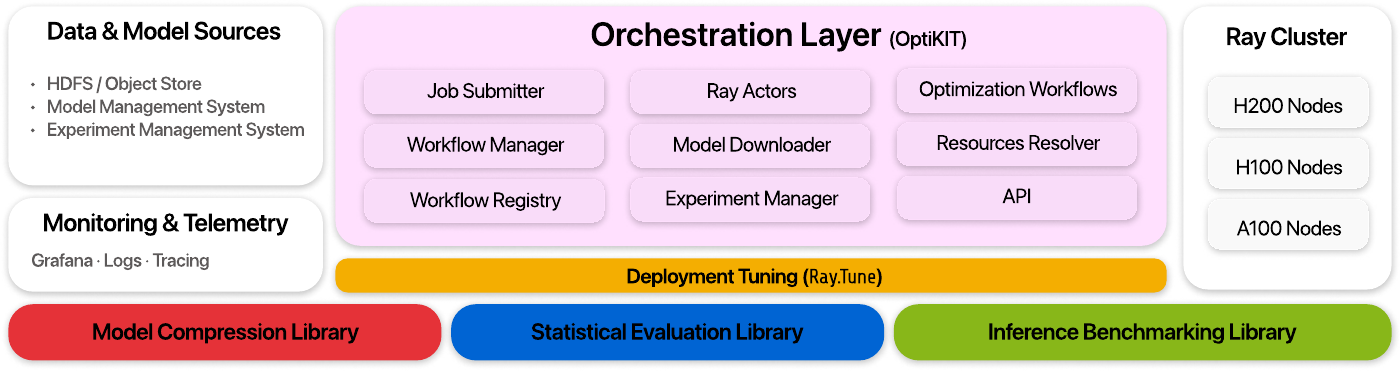}
\vspace{-10pt}
\caption{\textbf{\textsc{OptiKIT} system architecture.} The orchestration layer manages submission, resource allocation, and experiment tracking via Ray Actors, integrating with external data/model sources and heterogeneous GPU clusters. Supporting libraries provide extensible optimization capabilities; monitoring ensures observability.}
\label{fig:architecture}
\end{figure*}

\section{System Design and Architecture}
\label{sec:system_design}

\subsection{Design Philosophy}

\paragraph{Motivation}

\textsc{OptiKIT} follows three principles addressing the core challenges of large-scale LLM optimization.

\textit{Automation:} End-to-end workflows for compression, calibration, and tuning are automated through declarative task definitions, ensuring reproducibility and consistency.

\textit{Resource Awareness:} Heterogeneous resources are orchestrated according to each stage’s compute and data characteristics, maximizing utilization and minimizing overhead.

\textit{Interoperability:} Standardized interfaces connect to existing registries, data sources, and experiment tracking systems, enabling seamless integration into enterprise infrastructure.

\paragraph{Abstractions}
\textsc{OptiKIT} represents each optimization operation as a single \textit{process} (Figure \ref{fig:main_fig}), consisting of ordered stages that collectively form a streamlined flow:

\begin{enumerate}[noitemsep, topsep=0pt]
    \item \textbf{Fetch:} \label{item: fetch}
    Retrieve the target model and, if provided, calibration data from remote storage to local workspace.
    \item \textbf{Model Compression:} Parallelize quantization trials to capture variance in calibration data sampling.
    \item \textbf{Statistical Evaluation:} Evaluate quantized models to measure accuracy and potential quality degradation.
    \item \textbf{Inference Benchmarking:} Measure serving performance of the evaluated models under controlled load.
    \item \textbf{Deployment Tuning:} Optimize runtime parameters such as parallelism, batch size, and context window.
    \item \textbf{Upload:} \label{item: upload}
    Store the optimized model, associated metrics, and metadata to centralized tracking repositories.
\end{enumerate}

Flows are defined declaratively, with \textsc{OptiKIT} automatically mapping stages to actor pools and resource allocations at runtime to ensure reproducible, auditable execution.

\paragraph{Design Rationale}
Our architecture explicitly trades algorithmic novelty for system reliability, failure isolation, and automation. Rather than inventing new quantization mathematics, \textsc{OptiKIT} relies on standard search heuristics (e.g., TPE, exponential bracketing) and off-the-shelf post-training quantization. This approach allows us to solve the primary enterprise bottleneck: operationalizing the workflow. For example, as shown in our abstractions, we intentionally separate statistical evaluation from benchmarking. This is necessary because quality checks require variable-length I/O for correctness, whereas SLO certification requires tightly controlled, fixed traffic profiles.

\subsection{Architecture Components}

\textsc{OptiKIT} is structured as a distributed Python SDK built on Ray \cite{moritz2018ray}, organized into three fundamental architectural layers that provide clear separation of concerns and enable flexible, scalable optimization workflows.

\paragraph{Actor-Based Execution Layer} The foundation layer consists of specialized Ray actors that handle specific optimization tasks. Each actor type encapsulates the logic for a particular operation and manages its own computational resources. Actors can be dynamically created with appropriate GPU/CPU allocations, scaled horizontally across the cluster, and terminated to free resources. This design enables fault isolation where individual actor failures don't compromise entire jobs, and efficient resource utilization through fine-grained allocation per computational profile.

\paragraph{Flow Composition Layer}
Flows implement the \texttt{BaseFlow} contract and compose low-level actors into executable pipelines. A flow is responsible for: instantiating and sizing actor pools for each stage, mapping declarative resource hints to concrete GPU/CPU allocations, queuing trial work and load-balancing it across available actors, and coordinating deterministic teardown to reclaim resources between stages. 

We intentionally keep actor pools stage-isolated and ephemeral: compression, statistical evaluation, and benchmarking have different compute/memory footprints and failure modes, so per-stage pools allow tailored sizing and stronger fault isolation. Deterministic teardown also resets GPU state between stages, improving run-to-run consistency; the trade-off is barrier-induced idle time under stragglers, discussed in Section~\ref{sec:conclusion}. Failure handling is explicit: transient actor errors trigger bounded retries, while persistent trial failures are recorded and excluded from further stages. 

Each flow is bound to a versioned Docker image that encapsulates its runtime environment and dependencies, ensuring reproducibility and isolation across releases. Flows are registered via a Flow Registry, enabling teams to add new workflows without touching core runtime code. Each flow emits an archive of trial metadata, metrics, and artifacts to support reproducibility and post-hoc analysis.

\paragraph{Submission Engine Layer}
The submission engine manages job validation, packaging, and distributed execution. 
It converts high-level job specifications into executable configurations, validates them against \texttt{Pydantic} schemas, and bundles required artifacts—including the full \textsc{OptiKIT} runtime—into a self-contained package for deployment. 
The engine coordinates authentication, resource allocation, and container orchestration through existing enterprise schedulers, providing a uniform interface for both local and remote execution.

\section{Core Subsystems}

\textsc{OptiKIT} integrates several specialized subsystems that provide distinct optimization capabilities while maintaining seamless interoperability through standardized interfaces.

\subsection{Optimizer: Universal Compression Framework}

The \textit{Optimizer} \sethlcolor{myred}\hl{subsystem} (Figures \ref{fig:main_fig}, \ref{fig:architecture}) serves as \textsc{OptiKIT}’s universal engine for model compression \cite{zhu2024survey, wang2024model, zhou2024survey2}, providing a consistent interface for applying diverse optimization techniques across heterogeneous inference backends. Its backend-extensible design abstracts away engine-specific APIs behind a narrow contract, enabling portable compression workflows when adapters are provided for a given backend.

\paragraph{Backend-Extensible Architecture} The \textit{Optimizer} defines a standardized \textit{Optimization-Backend} interface that decouples \textsc{OptiKIT} flows from any specific compression library or inference engine. The contract exposes a small set of primitives (e.g., model load/save, strategy configuration, optimization execution), enabling alternative implementations to be swapped without changing workflow logic.
In this paper, we instantiate the interface with our current production backend (integrating vLLM-compatible compression routines~\cite{kwon2023efficient,llmcompressor2024}); supporting additional backends (e.g., TensorRT-LLM~\cite{tensorrtllm2023}) requires implementing the same contract. Our evaluation uses vLLM for serving and benchmarking; other serving backends are not evaluated here.

\paragraph{Recipe-Based Configuration System}
The \textit{Optimizer} introduces a \textit{recipe-based configuration} paradigm that transforms model compression from ad-hoc tuning into a structured, declarative workflow. A \textit{recipe} encodes a complete compression strategy—quantization scheme, calibration requirements, and layer-selection policy—into a reusable specification that captures domain heuristics such as layer exclusions and dataset size. Current recipes include:
\begin{itemize}[noitemsep, topsep=0pt]
    \item \texttt{int\_w8a8} and \texttt{int\_w4a16} — integer quantization recipes based on GPTQ~\cite{frantar2023gptq} and SmoothQuant~\cite{xiao2023smoothquant}, representing robust post-training quantization and activation balancing.
    \item \texttt{fp8\_dynamic} — a mixed-precision recipe derived from RTN~\cite{micikevicius2022fp8}, suitable for layers sensitive to integer quantization.
\end{itemize}

This abstraction standardizes compression workflows across models and tasks while remaining easily extensible. New recipes can be registered to integrate emerging quantization methods or custom heuristics without modifying core optimization logic.

\paragraph{Calibration Data Sampling} To support data-aware quantization \cite{xiao2023smoothquant,frantar2023gptq}, the \textit{Optimizer} includes a modular sampling pipeline for calibration dataset preparation. As established in prior work \cite{williams2023impact, zhang2025selectq}, selecting the correct calibration data affects the performance of the quantized model. Accordingly, the developed module supports multiple data calibration pipelines. These range from uniform random sampling to more advanced strategies such as length-weighted and token-statistics–stratified sampling. The goal is to account for variations in data distribution, dataset composition, and token-level characteristics. The module also provides hooks for easy extension with new strategies. Its flexible design further enables future adaptive calibration, where sampling dynamically adjusts to quantization performance.


\paragraph{Automated Optimization Workflow} The \textit{Optimizer} automates the full compression lifecycle, including model loading, calibration preprocessing, quantization, and model serialization. Calibration sample counts are derived directly from recipe specifications, and the pipeline orchestrates execution end-to-end, enabling fully automated, reproducible compression across backends.

\subsection{StatEval: LLM Statistical Evaluation library}
\label{sec:stateval}

The \textit{StatEval} \sethlcolor{myblue}\hl{package} (Figures \ref{fig:main_fig}, \ref{fig:architecture})
is a core component of the optimization framework, handling statistical evaluation of model performance across multiple inference backends. 

\paragraph{Design and Integration} \textit{StatEval} is built with a modular architecture that cleanly separates model handling from evaluation logic. The package is designed for seamless integration with internal eBay infrastructure while remaining easily adaptable to open-source and external environments. It supports two primary runtime interfaces: local vLLM for offline evaluation and OpenAI-style HTTP endpoints for online inference. Engines that expose OpenAI-compatible APIs (e.g., vLLM or SGLang \cite{zheng2024sglangefficientexecutionstructured}) can be integrated with minimal changes; other engines require adapters. In this paper, we use vLLM for offline evaluation.

Within the workflow, \textit{StatEval} acts as a quality-feasibility gate before performance search. We keep this stage separate from benchmarking because quality runs prioritize task-faithful prompts and scoring pipelines, whereas serving certification requires controlled arrival processes; conflating them is methodologically nonsensical, since the measured performance would primarily reflect benchmark-harness characteristics rather than deployment traffic and SLO behavior (see Section~\ref{sec:benchmarker}).


\paragraph{Supported Benchmarks}
\textit{StatEval} is an internal package tailored for e-commerce–specific model evaluation. The package includes open-source benchmarks to enable standardized and comparable assessment: \textit{GSM8K} \cite{gsm}, \textit{IFEval} \cite{ifeval}, \textit{Do-Not-Answer} \cite{donotanswer}. The selected benchmarks cover three core LLM capabilities—reasoning, instruction-following, and safety—essential for both general use-cases and e-commerce applications. Additionally, we develop in-house benchmarks that are a core component of the system. They act as critical proxies of e-commerce production metrics for rapid experimentation cycles. These benchmarks are proprietary and excluded from this study.



\subsection{Benchmarker: Performance Testing Tool}
\label{sec:benchmarker}

The \emph{Benchmarker} \sethlcolor{mygreen}\hl{library} (Figures \ref{fig:main_fig}, \ref{fig:architecture}) quantifies the performance capabilities of an optimized model while ensuring compliance with predefined SLOs such as end-to-end latency and time per output token. Its purpose is to determine the maximum sustainable request rate that maintains target SLOs under a fixed serving profile $\Pi$. Because $\Pi$ encodes request mix and input/output token-length characteristics, certified throughput is workload- and SLO-dependent rather than model-intrinsic. 

The \emph{Benchmarker} executes controlled load experiments on the optimized model within the Ray cluster, monitoring fine-grained telemetry via integrated tracing and metrics pipelines. These metrics serve as a critical interface between optimization workflows and deployment configurations, i.e., guaranteeing the required performance envelope for production rollout. 

We structure SLO-gated benchmarking as a low-cost closed-loop infeasibility gate followed by open-loop certification: the first rejects configurations that fail even without queueing pressure, while the second measures sustainable throughput as batching and queueing emerge under $\Pi$. Moreover, by standardizing the SLO-driven benchmarking process, we ensure comparability across experiments, hardware classes, and compression strategies. The full procedure is summarized in Algorithm~\ref{alg:benchmarker-sweep}, which outlines the iterative search and decision logic governing the sweep. Below, we dive into core algorithmic components.

\paragraph{Steady-State Regression} \label{sec: stead_state}
Assessing whether the system has reached steady state involves modeling the relationship between \emph{request arrivals} and \emph{completions} as a linear process.
Compared to using a single, statically configured warmup time, this regression-based criterion provides a simple robustness check: it can detect when a trial is still ``settling'' (e.g., delayed initialization, lazy CUDA graph capture, or transient stalls) even if the run has already passed a fixed warmup period.
In our implementation, we still apply a short fixed warmup to amortize one-time initialization overheads, and then use the regression signal during the measurement window as the accept/reject gate for stability.
Specifically, we fit a regression of the form

\begin{figure}[t]
\centering
\includegraphics[width=0.38\textwidth]{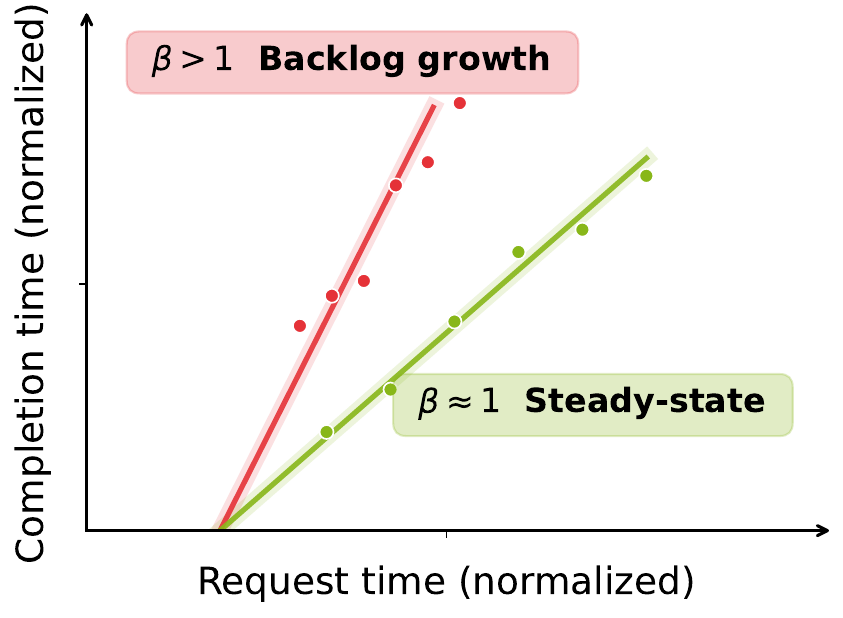}
\vspace{-5pt}
\caption{\textbf{Regression diagnostics for a Benchmarker trial.} A fitted slope $\beta \approx 1$ (\textit{green}) indicates steady-state operation queuing while $\beta > 1$ (\textit{red}) denotes an overload regime.}
\label{fig:regression}
\end{figure}

\begin{equation} \label{eq: 1}
    r_i = \alpha + \beta\,c_i + \varepsilon_i,
\end{equation}

where $\alpha$ is a fixed overhead, $r_i$ denotes the arrival timestamp and $c_i$ the completion timestamp of request~$i$ (both normalized relative to the start time), and $\varepsilon_i$ captures residual noise.
The estimated slope~$\beta$ acts as a compact indicator of system equilibrium:
\begin{itemize}[noitemsep, topsep=0pt]
    \item $\beta \approx 1$: \textbf{Steady-state} --- completions keep pace with arrivals, and the queue length remains stable.
    \item $\beta > 1$: \textbf{Overloaded} --- arrivals faster than completions, leading to backlog accumulation and latency inflation.
\end{itemize}
A trial is considered stable if $\lvert \beta - 1\rvert \le \tau_\beta$, where $\tau_\beta$ is a small tolerance (typically 0.02--0.05 depending on noise and sampling granularity).
In addition to the slope, the regression intercept and correlation coefficient are logged as part of the diagnostic record, providing visibility into drift patterns and fit quality. Figure~\ref{fig:regression} illustrates a typical diagnostic plot.
A fitted slope near 1 indicates arrivals and completions are matched. Deviations from 1 reveal rate imbalance and queue growth, enabling the Benchmarker to find the highest sustainable load before instability.

Beyond a short fixed warmup, we apply a regression-based stability check during the measurement window to handle variable initialization (e.g., lazy CUDA graph capture, dynamic batching warmup). This approach:
\begin{itemize}[leftmargin=1.5em,itemsep=0pt]
    \item \textbf{Detects stability violations:} By monitoring $\beta \approx 1$ (Eq.~\ref{eq: 1}), we reject runs corrupted by insufficient warmup or runtime stalls.
    \item \textbf{Ensures data quality:} A tight tolerance ($\tau_\beta=0.02$) prioritizes clean measurements. Our 900s trials ($\sim$thousands of requests) ensure statistical fit quality.
\end{itemize}

\paragraph{Exponential Search}
\emph{Benchmarker} explores the feasible operating region through an adaptive sweep over candidate request rates.
Each trial is executed at a fixed request rate and evaluated against both the SLO criteria and the steady-state condition described above.

Rates that satisfy all constraints are marked as \emph{passing}, while those that violate any constraint are marked as \emph{failing}.
After each evaluation, the next test rate is selected using a \emph{bounded search heuristic} that incrementally narrows the feasible region. We use exponential bracketing to quickly localize the feasibility boundary, then midpoint refinement to approximate the maximum SLO-feasible rate with logarithmic trial growth; this is more sample-efficient than dense fixed-grid sweeps when each trial is expensive.

If this baseline already violates the specified SLOs, the configuration is deemed
infeasible and the sweep terminates early.
The final output comprises the highest passing rate---the system’s sustainable throughput under SLO compliance---together with a complete archive of all tested rates, stability diagnostics, and latency statistics.
This archive supports reproducibility, post-hoc analysis, and cross-hardware comparability across optimization trials.

We employ a two-phase approach to efficiently certify SLO compliance:

\begin{itemize}[leftmargin=1.5em,itemsep=0pt]
    \item \textbf{Phase 1 (Closed-loop infeasibility gate):} A low-cost check that rejects configurations failing the latency SLO even under zero queueing pressure. This eliminates obviously infeasible configurations without expensive rate sweeps.

    \item \textbf{Phase 2 (Open-loop certification):} An asynchronous constant-rate trial that measures sustainable throughput under a fixed serving profile $\Pi$. By fixing the arrival process, we inherently capture the emergent batching, queue dynamics, and stability as request rate $r$ increases---without analytically "correcting" for batching effects.
\end{itemize}

This structure bounds search cost via early rejection while ensuring that certified configurations can sustain their claimed throughput under realistic queueing and batching behavior.

Algorithm~\ref{alg:benchmarker-sweep} uses an adaptive search to efficiently pinpoint the system's saturation boundary. It implements exponential doubling ($2r$) of the request rate upon success to rapidly explore capacity limits, and dynamically switches to midpoint halving (bisection) upon failure to refine the exact threshold. This achieves logarithmic complexity, avoiding the prohibitive compute costs of exhaustive grid search while preventing premature convergence.

\begin{algorithm}[!t]
\caption{\emph{Benchmarker} Sweep (closed-loop infeasibility gate + open-loop certification under fixed profile $\Pi$)}\label{alg:benchmarker-sweep}
\begin{algorithmic}[1]
\REQUIRE Serving profile $\Pi = \langle$input/output length distributions, request mix, think time, \dots$\rangle$ \\
(Optional) SLOs $\mathcal{S}=\{\}$, Error margins $\mathcal{E}=\{\}$ \\
(Defaults) initial rate $r_0$, budget $N$, threshold $\mathcal{T}$
\STATE $\textit{best}\gets\text{none}$
\IF{SLOs provided}
  \STATE Run a \textbf{synchronous} \textit{closed-loop} trial under $\Pi$
  \IF{$\exists s \in \mathcal{S} \sim \mathcal{E}$ violated}
    \STATE \textbf{return} $\langle \mathrm{status}:\mathrm{INFEASIBLE}, \mathrm{rate}:0.0 \rangle$
  \ELSE
    \STATE $\mathcal{LB} \gets \mathbb{E}[\mathrm{latency}^{-1}]$ (lower bound rate)
  \ENDIF
\ENDIF
\STATE $r \gets r_0$ 
\WHILE{not converged \AND no.\ trials $\leq N$}
  \STATE Run an \textbf{asynchronous} \textit{open-loop} trial at $r$ under $\Pi$
  \IF{$\forall s \in \mathcal{S} \sim \mathcal{E}$ passed \AND queuing \textit{steady-state}} 
     \STATE \textit{best} $\gets r$
     \STATE $r \gets 2r$ (exponential doubling)
  \ELSE
     \STATE $r\gets \frac{(\mathcal{LB} + r)}{2}$ (midpoint halving)
  \ENDIF
  \IF{$|best - r| \leq \mathcal{T}$}
    \STATE set converged
  \ENDIF
\ENDWHILE
\STATE \textbf{return} $\langle \mathrm{status}:\mathrm{FEASIBLE}, \mathrm{rate}:best \rangle$
\end{algorithmic}
\end{algorithm}

\paragraph{Traffic Pattern.}
The open-loop trials use a Poisson arrival process with mean rate $r$ requests/second. Each request draws input/output lengths from the workload-specific distribution defined in $\Pi$ (see Section~\ref{sec:experiments} for per-workload profiles). This asynchronous pattern allows queue buildup and realistic batching to emerge naturally as $r$ approaches the saturation point.

\begin{algorithm}[!th]
\caption{Quantization with Tuning Flow (trial-parallel, actor-pool based)}
\label{alg:quantize-tune}
\begin{algorithmic}[1]
\REQUIRE Model $\mathcal{M}$, dataset $\mathcal{D}$ (optional), number of trials $N_{\text{trials}}$, resource budget $R$
\STATE \textbf{Fetch} $\mathcal{M}$ and $\mathcal{D}$ from remote storage; store locally
\STATE Sample $N_{\text{trials}}$ distinct calibration subsets $\{\mathcal{C}_i\}_{i=1}^{N_{\text{trials}}}$
\STATE \textbf{results} $\gets \varnothing$
\STATE \textbf{create} quantization actor pool sized to $R$
\FORALL{$i \in \{1,\dots,N_{\text{trials}}\}$ \textbf{in parallel}}
    \STATE Apply quantization recipe to $\mathcal{M}$ with calibration $\mathcal{C}_i$; produce compressed model $q_i$
    \STATE Attach metadata (seed, recipe, path) and append $\langle \mathcal{C}_i,q_i\rangle$ to \textbf{results}
\ENDFOR
\STATE \textbf{destroy} quantization actor pool \COMMENT{free GPUs and reset distributed state}
\STATE \textbf{create} evaluation actor pool sized to $R$
\FORALL{each compressed model $q$ in \textbf{results} \textbf{in parallel}}
    \STATE Run statistical evaluation on $q$; attach quality metrics to its record
\ENDFOR
\STATE \textbf{destroy} evaluation actor pool
\STATE Let $\mathcal{S}$ be successful candidates
\IF{$\mathcal{S} = \varnothing$}
    \STATE \textbf{return} failure status and archive
\ENDIF
\STATE Select representative quantized model $q^\ast \in \mathcal{S}$
\STATE \textbf{create} benchmarking actor pool sized to $R$
\STATE Benchmark $q^\ast$ (and optionally full-precision baseline) to collect runtime + stability metrics
\STATE \textbf{destroy} benchmarking actor pool
\STATE Create single CPU tuning orchestrator
\STATE Build tuning search space (tensor parallel sizes, max\_num\_seqs, max\_num\_batched\_tokens, \dots)
\FORALL{configuration $c$ proposed by tuner (Ray Tune / Optuna) \textbf{in parallel or sequential as resources permit}}
    \STATE Instantiate benchmark job for $(q^\ast, c)$
    \STATE Measure metrics (throughput, normalized request rate, pass\_slo, etc.)
    \STATE Report metrics back to tuner
\ENDFOR
\STATE Destroy tuning orchestrator
\STATE Persist: quantized model $q^\ast$, best tuning configuration $c^\ast$, metrics, and trial archive to EMS / model registry
\STATE \textbf{return} $\{q^\ast, c^\ast, \text{trial archive}\}$
\end{algorithmic}
\end{algorithm}

\subsection{Tuner: Automated Hyperparameter Optimization}

The tuning \sethlcolor{myyellow}\hl{stage} (Figures \ref{fig:main_fig}, \ref{fig:architecture}) optimizes the runtime configuration of the quantized model to maximize inference throughput while maintaining compliance with SLOs.
Rather than modifying model weights, it searches over inference engine runtime parameters that control parallelism, batching, and context allocation.
Quality constraints are enforced by the preceding StatEval gate; incorporating accuracy explicitly into the tuning loop is straightforward by adding quality metrics to the reported objective and treating them as additional constraints (or Pareto dimensions) alongside latency.
Similarly, decision variables such as speculative decoding or MoE expert-parallel configuration can be surfaced as additional tuning dimensions whenever the serving backend exposes them; in this paper we focus on the core vLLM runtime knobs available in our deployment setting.
The \texttt{TunerActor} integrates with the \texttt{BenchmarkerActor} subsystem to evaluate each candidate configuration under realistic serving workloads. Every Ray Tune trial executes a complete benchmark evaluation, measuring throughput, latency, and SLO compliance.
\vspace{-10pt}
\paragraph{Optimization Objective} The optimization objective combines these metrics into a single scalar fitness function, defined as:

\begin{equation}
\text{fitness}(c) = \tfrac{\text{throughput}(c)}{\text{tensor\_parallel\_size}(c)} + \lambda \cdot \text{slo\_penalty}(c)
\end{equation}

where $c$ denotes a candidate configuration. Throughput is normalized per GPU to ensure fair comparison across different parallelization strategies, and $\lambda$ applies a large negative penalty for SLO violations (typically $\lambda = -1000$). This formulation guides the search toward configurations that sustain high per-GPU throughput while satisfying latency and stability requirements. Our formulation maximizes throughput subject to a strict latency SLO. This aligns with production goals: serving latency acts as a binary constraint, meaning that once the target is met, system efficiency is strictly governed by throughput maximization.
\vspace{-10pt}
\paragraph{Tuning Orchestration} The tuning process is orchestrated by a single \texttt{TunerActor}, which builds its parameter search space using the same input and output configuration applied during benchmarking of the quantized model. This ensures that the tuning trials explore serving parameters under identical workload conditions, preserving consistency in sequence lengths, token limits, and request patterns. Each tuning trial spawns a temporary \texttt{BenchmarkerActor}, which launches a vLLM server, generates synthetic request batches, and runs steady-state load tests to measure request rate and SLO pass ratio. 

The search explores key inference engine parameters that influence runtime efficiency:
\begin{itemize}[noitemsep, topsep=0pt]
    \item \textbf{KV-cache / max-context provisioning:} We tune the \texttt{maximum context size} (i.e., the KV-cache provisioning target) computed from user-specified input and output length requirements as $(input\_len + output\_len) \times 1.15$. Over-provisioning reserves additional KV memory and can reduce the feasible concurrency ceiling, while under-provisioning risks truncation or OOMs.
    \item \textbf{Parallelism\,Strategies:} The search space for \texttt{tensor} and \texttt{data} parallelism explores configurations such as $\{1, 2, 4, 8\}$ bounded by cluster resource availability and user-defined limits.
    \item \textbf{Batch\,Processing:} Other parameters such as \texttt{maximum concurrency} and \texttt{maximum token batch size} are tuned within user-configurable ranges or system defaults.
\end{itemize}

Ray Tune employs the Optuna \cite{akiba2019optuna} search algorithm, which implements Tree-structured Parzen Estimators (TPE) \cite{watanabe2023tree} for Bayesian optimization. This strategy models the objective landscape probabilistically and selects configurations that balance exploration of new regions with exploitation of known high-performing areas, improving sample efficiency compared to random or grid search. Each configuration is benchmarked using the same load generation and measurement logic as the performance stage, and metrics are reported back to Ray Tune. The best-performing configuration, its associated metrics, and the full archive of evaluated trials are stored with the quantization artifacts and uploaded at the final pipeline stage.


\subsection{Final Algorithm} \label{sec:final_algorithm}
Algorithm~\ref{alg:quantize-tune} integrates the full \textsc{OptiKIT} \textit{Quantization with Tuning Flow} (elaborating on Figure~\ref{fig:main_fig}) into a unified, resource-aware pipeline.
The flow executes as a sequence of distributed stages; computationally independent stages are parallelized across actor pools sized to available GPU/CPU resources.
Each actor processes one trial at a time, and results are synchronized before proceeding.
Actor pools are explicitly destroyed between stages to free GPU memory and reset distributed state, ensuring deterministic resource reclamation and reproducible runs.

%% file: sections/04_experiments.tex
\section{Experiments}
\label{sec:experiments}

\subsection{Experimental Setup}

For our experiments, we used NVIDIA H100 GPUs for both quantization and inference tuning tasks.
Each experiment was executed within the same environment to ensure consistency and comparability across models and configurations.
Through this setup, we evaluated both the statistical performance recovery of quantized models and the inference performance gains achieved through runtime tuning.
The end-to-end optimization runtime along with the GPU Hours for each evaluated model and stage are summarized in Figure~\ref{fig: total_optikit_cost} and in Paragraph \ref{par: optikit computational cost analysis}.

\subsection{Evaluated Models and Configurations}

We tested three open-source LLMs representative of different operational scales and latency requirements:
Qwen 2.5 7B Instruct~\cite{qwen2_5},
Mistral Small 3 24B Instruct~\cite{mistral_small}, and
Meta Llama 3.3 70B Instruct~\cite{grattafiori2024llama3herdmodels}.

For each model, we applied three quantization recipes available in \textsc{OptiKIT}:
Dynamic FP8, INT W8A8 (static-weight / dynamic-activation), and INT W4A16 (static-weight / high-precision activation).
For the INT-based configurations, we used the default calibration dataset from \cite{calibration_dataset}, performing five independent trials with 256 random calibration samples for W8A8 and 512 for W4A16 per trial.
The FP8 configuration required no calibration data and therefore exhibits no trial variance.

This setup enabled a direct comparison between quantization precision, calibration strategy, and runtime optimization under realistic production-style workloads.

\subsection{Statistical Performance}

We used \textsc{OptiKIT} to measure the impact of different quantization recipes on each model’s statistical performance.
Table~\ref{tab:statistical_performance} reports the best-performing trial result, the corresponding recovery ratio relative to the full-precision baseline, mean, standard deviation (STD) and relative standard deviation (RSD) across five trials. For GSM8K, we report 8-shot exact match in the Chain-of-Thought setting; for Do-Not-Answer, the harmless responses proportion; and for IFEval, the mean of prompt- and instruction-level accuracy, following Meta’s recipe~\cite{grattafiori2024llama3herdmodels}.

\begin{table}[!t]
\centering
\caption{\textbf{Example Inference Use-cases.} Representative eBay-derived inference use cases, with model scale, input/output token ratios, and corresponding latency SLOs.}
\vspace{-8pt}
\vskip 0.1in
\label{tab:usecases}
\renewcommand{\arraystretch}{1.15}
\resizebox{\columnwidth}{!}{%
\begin{tabular}{lcc}
\toprule
\textbf{Model} & \textbf{Input / Output (prefix)} & \textbf{Performance Objective} \\
\midrule
Llama 3 70B & 5000 / 500 & Throughput-oriented \\
Qwen 2.5 7B & 1200 / 80 & Latency p95 $\leq$ 500\,ms \\
Mistral Small 3 24B & 3000 / 200 (2000) & Latency p95 $\leq$ 1500\,ms \\
Mistral Small 3 24B & 1500 / 1500 (1000) & TTFT p50 $\leq$ 50\,ms, TPOT p50 $\leq$ 10\,ms \\
\bottomrule
\end{tabular}%
}
\vskip -0.1in
\end{table}

\begin{table*}[!ht]
\centering
\small
\caption{\textbf{Statistical performance across models.} Performance recovery and trial variance across quantization recipes, models, and benchmarks.}
\vspace{-8pt}
\vskip 0.1in
\label{tab:statistical_performance}
\renewcommand{\arraystretch}{1.15}
\begin{adjustbox}{width=\textwidth}
\begin{tabular}{cccccccccccl}
\toprule
\multicolumn{1}{c|}{} & \multicolumn{1}{c|}{\textbf{Full precision}} & \multicolumn{2}{c|}{\textbf{FP8 Dynamic}} & \multicolumn{4}{c|}{\textbf{INT W8A8}} & \multicolumn{4}{c}{\textbf{INT W4A16}} \\ \hline
\multicolumn{1}{c|}{\textbf{Task}} & \multicolumn{1}{c|}{\textbf{Result}} & \textbf{Result} & \multicolumn{1}{c|}{\textbf{Recovery}} & \textbf{Result} & \textbf{Recovery} & \textbf{Mean} & \multicolumn{1}{c|}{\textbf{STD (RSD \%)}} & \textbf{Result} & \textbf{Recovery} & \textbf{Mean} & \textbf{STD (RSD \%)} \\ \hline
\multicolumn{12}{c}{\textit{Qwen 2.5 7B Instruct}} \\ \hline
\multicolumn{1}{c|}{GSM8K} & \multicolumn{1}{c|}{0.826} & 0.818 & \multicolumn{1}{c|}{99.031\%} & 0.823 & 99.637\% & 0.821 & \multicolumn{1}{c|}{0.003 (0.365\%)} & 0.807 & 97.7\% & 0.811 & 0.005 (0.617\%) \\
\multicolumn{1}{c|}{IFEval} & \multicolumn{1}{c|}{0.773} & 0.758 & \multicolumn{1}{c|}{98.06\%} & 0.767 & 99.224\% & 0.764 & \multicolumn{1}{c|}{0.003 (0.393\%)} & 0.795 & 102.846\% & 0.761 & 0.02 (2.628\%) \\
\multicolumn{1}{c|}{Do-Not-Answer} & \multicolumn{1}{c|}{0.970} & 0.972 & \multicolumn{1}{c|}{100.206\%} & 0.973 & 100.309\% & 0.973 & \multicolumn{1}{c|}{0.001 (0.103\%)} & 0.967 & 99.691\% & 0.967 & 0.002 (0.207\%) \\
\hline
\multicolumn{12}{c}{\textit{Mistral Small 3 24B Instruct}} \\ \hline
\multicolumn{1}{c|}{GSM8K} & \multicolumn{1}{c|}{0.868} & 0.864 & \multicolumn{1}{c|}{99.539\%} & 0.879 & 101.267\% & 0.876 & \multicolumn{1}{c|}{0.007 (0.799\%)} & 0.873 & 100.576\% & 0.862 & 0.008 (0.928\%) \\
\multicolumn{1}{c|}{IFEval} & \multicolumn{1}{c|}{0.784} & 0.777 & \multicolumn{1}{c|}{99.107\%} & 0.733 & 93.495\% & 0.718 & \multicolumn{1}{c|}{0.01 (1.393\%)} & 0.780 & 99.49\% & 0.776 & 0.005 (0.644\%) \\
\multicolumn{1}{c|}{Do-Not-Answer} & \multicolumn{1}{c|}{0.945} & 0.946 & \multicolumn{1}{c|}{100.106\%} & 0.936 & 99.048\% & 0.941 & \multicolumn{1}{c|}{0.004 (0.425\%)} & 0.946 & 100.106\% & 0.952 & 0.004 (0.42\%) \\
\hline
\multicolumn{12}{c}{\textit{Llama 3.3 70B Instruct}} \\ \hline
\multicolumn{1}{c|}{GSM8K} & \multicolumn{1}{c|}{0.914} & 0.915 & \multicolumn{1}{c|}{100.109\%} & 0.909 & 100.219\% & 0.909 & \multicolumn{1}{c|}{0.005 (0.55\%)} & 0.907 & 99.234\% & 0.908 & 0.001 (0.11\%) \\
\multicolumn{1}{c|}{IFEval} & \multicolumn{1}{c|}{0.912} & 0.920 & \multicolumn{1}{c|}{100.877\%} & 0.912 & 101.206\% & 0.915 & \multicolumn{1}{c|}{0.005 (0.546\%)} & 0.915 & 100.329\% & 0.917 & 0.004 (0.436\%) \\
\multicolumn{1}{c|}{Do-Not-Answer} & \multicolumn{1}{c|}{0.995} & 0.949 & \multicolumn{1}{c|}{95.377\%} & 0.948 & 95.578\% & 0.949 & \multicolumn{1}{c|}{0.003 (0.316\%)} & 0.951 & 95.578\% & 0.946 & 0.003 (0.317\%) \\
\bottomrule
\end{tabular}
\end{adjustbox}
\vskip -0.1in
\end{table*}

\begin{table}[t]
\centering
\small
\renewcommand{\arraystretch}{1.2}
\caption{\textbf{Qwen 2.5 7B Instruct \emph{without} vs. \emph{with} determinism.} We report results for Qwen 2.5 7B Instruct (100 runs per task) \emph{without} vs. \emph{with} vLLM deterministic setting. The deterministic configuration yields nearly identical results across trials.}
\vspace{-8pt}
\vskip 0.1in
\begin{tabular}{l cccc}
\toprule
\multirow{2}{*}{\textbf{Tasks}} & \multicolumn{4}{c}{\textbf{Statistics}} \\
 & Min & Max & Mean & STD (RSD \%) \\
\midrule
\multicolumn{5}{c}{\textit{Non-deterministic} setting} \\
 GSM8K & 0.818 & 0.829 & 0.824 & 0.002 (0.243\%) \\
 IFEval & 0.754 & 0.781 & 0.767 & 0.005 (0.652\%) \\
 \midrule
\multicolumn{5}{c}{\textit{Deterministic} setting} \\
GSM8K & 0.822 & 0.822 & 0.822 & 0.000 (0.0\%) \\
IFEval & 0.771 & 0.775 & 0.773 & 0.001 (0.129\%) \\
\bottomrule
\end{tabular}
\vspace{-5pt}
\label{tab:determinism_combined}
\vskip -0.1in
\end{table}

Across all evaluated tasks, both FP8 Dynamic and INT W8A8 quantization achieved near full-precision performance, with average recovery rates exceeding 99\%. For Qwen 2.5 7B, performance degradation was minimal—typically below 0.5\%, and in some cases, quantized models slightly surpassed full-precision baselines, indicating robustness to reduced precision. Similarly, Mistral Small 3 24B retained strong accuracy, with FP8 and INT8 models maintaining within 1\% of the original results on average. However, INT8 showed higher variability across tasks (RSD up to 1.4\%), reflecting task-dependent sensitivity. Mistral exhibited the greatest degradation on the IFEval task (93.5\% recovery), showing reduced ability to follow multiple instructions simultaneously compared to the full-precision counterpart. For Llama 3.3 70B, quantization maintained near-identical performance to full precision on GSM8K and IFEval (usually with recovery greater than 100\%), while Do-Not-Answer exhibited a modest reduction to around 95\% recovery.

Overall, FP8 and INT8 quantization effectively preserved model performance with minimal loss, whereas INT4, while viable in some cases, exhibited inconsistent behavior and greater sensitivity to task characteristics.

\paragraph{Run-to-run Consistency and Determinism}

Production settings place particular emphasis on run-to-run consistency and deterministic behavior, since evaluation results often drive automated model selection decisions. However, achieving strict determinism in LLM inference is challenging due to non-determinism in model execution and CUDA kernels.
Table \ref{tab:determinism_combined} presents results from 100 runs, illustrating variability under default vLLM settings versus deterministic mode \cite{kwon2023efficient}. Although disabling multiprocessing yields nearly deterministic results, it prevents deloading VRAM in one Python interpreter session and is only applicable to offline inference, thus imposing practical limitations.

When full determinism cannot be achieved, it is crucial to assess whether observed differences are statistically significant—especially when evaluation results guide automatic model selection, where random variation may lead to incorrect conclusions. Addressing this requires controlled evaluation protocols and statistically grounded comparison methods to ensure robust, reliable assessment, which will be a focus of future development.


\subsection{Inference Performance}

We next examined the impact of quantization and runtime tuning on inference efficiency.
Each workload configuration in Table~\ref{tab:usecases} represents a characteristic operational regime—varying in input–output token ratios, SLOs, and model scale—to reflect eBay’s production inference patterns.

For each workload, we performed a controlled benchmarking study to disentangle the contributions of quantization and runtime tuning.
The baseline used the FP16 model with default vLLM parameters.
The quantization-only setup applied model compression while keeping vLLM defaults, isolating quantization effects.
The tuning-only setup optimized the FP16 runtime configuration using \textsc{OptiKIT}’s deployment tuner.
Finally, the end-to-end configuration combined both quantization and tuned vLLM parameters to assess their joint impact.
Each tuning study employed TPE optimization over 30 trials, jointly searching \texttt{max\_num\_seqs}, \texttt{max\_num\_batched\_tokens}, and \texttt{tensor\_parallel\_size}.

\begin{table*}[!t]
\centering
\small
\caption{\textbf{Normalized per-GPU throughput and improvement vs.\ FP16 baseline.} Normalized throughput per GPU (SLO-compliant); improvements as multiplicative factors vs.\ baseline. ($^{*}$) marks SLOs not met without optimization.}
\vspace{-8pt}
\vskip 0.1in
\label{tab:results}
\renewcommand{\arraystretch}{1.1}
\begin{tabularx}{\textwidth}{@{} l
    >{\centering\arraybackslash}X
    >{\centering\arraybackslash}X >{\centering\arraybackslash}X
    >{\centering\arraybackslash}X >{\centering\arraybackslash}X
    >{\centering\arraybackslash}X >{\centering\arraybackslash}X @{}}
\toprule
\multirow{2}{*}{\textbf{Model}}
  & \multirow{2}{*}{\textbf{Baseline}}
  & \multicolumn{2}{c}{\textbf{Quantization only}}
  & \multicolumn{2}{c}{\textbf{Tuning only}}
  & \multicolumn{2}{c}{\textbf{Quantization + Tuning}} \\
\cmidrule(lr){3-4}\cmidrule(lr){5-6}\cmidrule(lr){7-8}
  &
  & \textbf{Norm. TPS} & \textbf{Gain}
  & \textbf{Norm. TPS} & \textbf{Gain}
  & \textbf{Norm. TPS} & \textbf{Gain} \\
\midrule
Qwen 2.5 7B
  & 3.52$^{*}$
  & 5.96 & 1.69$\times$
  & 6.79 & 1.93$\times$
  & \textbf{7.50} & \textbf{2.13$\times$} \\
Mistral 24B (Latency p95)
  & 0.604$^{*}$
  & 1.732 & 2.87$\times$
  & 0.937 & 1.55$\times$
  & \textbf{1.734} & \textbf{2.87$\times$} \\
Mistral 24B (TTFT \& TPOT p50)
  & 0.562$^{*}$
  & 0.562 & 1$\times$
  & 0.750 & 1.33$\times$
  & \textbf{0.875} & \textbf{1.55$\times$} \\
Llama 3 70B
  & 0.468$^{*}$
  & \textbf{0.593} & \textbf{1.26$\times$}
  & 0.468 & 1$\times$
  & 0.585 & 1.25$\times$ \\
\bottomrule
\end{tabularx}
\vspace{2pt}
\raggedright
\footnotesize $^{*}$SLOs not met without either quantization or tuning at the indicated tensor-parallel levels (TP=1, 2).
\vskip -0.1in
\end{table*}


\begin{figure*}[htbp]
    \centering
    \begin{subfigure}[t]{0.48\textwidth}
        \centering
        \includegraphics[width=\textwidth]{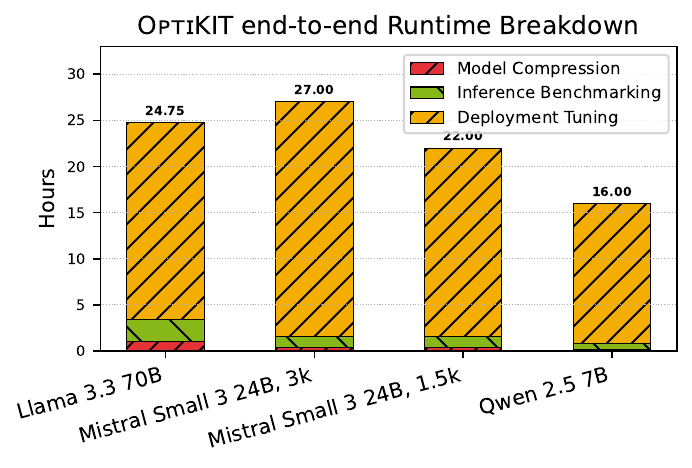} 
        \caption{\textbf{Wall-clock runtime breakdown.} Total optimization flow time per model and stage.}
        \label{fig: end-to-end-runtime}
    \end{subfigure}\hfill
    \begin{subfigure}[t]{0.48\textwidth}
        \centering
        \includegraphics[width=\textwidth]{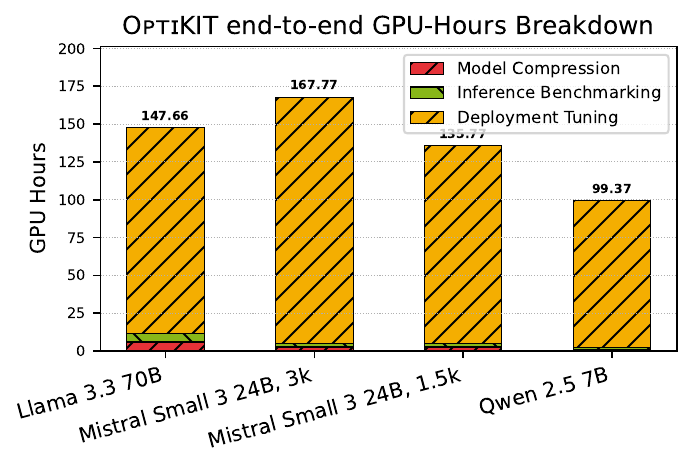} 
        \caption{\textbf{GPU-hours breakdown.} Total GPU-hours consumed across all stages. Considering parallelization, GPUs used per trial and scheduling issues.}
        \label{fig: gpu_hours_end-to-end-runtime}
    \end{subfigure}

    \vspace{-4pt}
    \caption{\textbf{\textsc{OptiKIT} computational cost analysis.} (a) shows wall-clock runtime, while (b) presents total GPU-hours consumed. In both plots we report the breakdown per single Stage, Model Compression, Inference Benchmarking, and Deployment Tuning, with \textit{StatEval} having negligible time ($\leq 0.15\%$) and therefore excluded. We report two Mistral Small 3 24B input/output scenarios as described in Table \ref{tab:usecases}.}
    \label{fig: total_optikit_cost}
\end{figure*}

The values in Table~\ref{tab:results} represent normalized per-GPU throughput for configurations that meet their respective SLOs.
We treat SLO compliance as a hard deployment constraint and optimize within the feasible set, rather than trading off SLO violations for higher raw throughput.
This normalization enables direct comparison across tensor-parallel regimes that may use different GPU counts and highlights the most cost-effective configuration—i.e., the setup yielding the highest SLO-compliant throughput per GPU.
Notably, for Qwen, SLOs were not met without either quantization or tuning at TP=1, while for Mistral, SLOs were not satisfied under TP=1 or TP=2 in the absence of these optimizations.

\paragraph{\textsc{OptiKIT} Computational Cost Analysis}
\label{par: optikit computational cost analysis}
In Figure \ref{fig: total_optikit_cost} we report, side by side, the total wall-clock \textsc{OptiKIT} runtime (\ref{fig: end-to-end-runtime}) and the total GPU-hours (\ref{fig: gpu_hours_end-to-end-runtime}).
On the left, Figure \ref{fig: end-to-end-runtime} reports the total time it takes to run \textsc{OptiKIT} per stage, regardless of the number of GPUs utilized. On the right, Figure \ref{fig: gpu_hours_end-to-end-runtime} instead breaks down GPU-hours, accounting for parallelization, the number of GPUs used per trial, and computational overheads caused by scheduling inefficiencies.
Deployment tuning constitutes the primary overhead, accounting for an average of $70\%$ of the total runtime. In contrast, statistical evaluation requires a negligible fraction of the overall time ($\leq 0.15\%$).

%% file: sections/05_conclusion.tex
\section{Discussion \& Insights}
\label{sec:conclusion}




\paragraph{Generalization of Quantization Quality}
Our results indicate that automated quantization with a generic calibration dataset \cite{calibration_dataset} achieves stable and robust, production-ready quality without expert supervision (Table \ref{tab:statistical_performance}). This successful generalization, however, raises new questions about its boundaries. It is unclear if this robustness would hold after domain-specific fine-tuning (e.g., LoRA), or if domain-aligned calibration data would become necessary. Furthermore, the impact of short-context calibration on long-context task fidelity, which was not evaluated, remains an open research question \cite{paglieri2024outliers}.


\paragraph{Tuning shines when SLOs are tight} With strict SLOs (latency p95 and TTFT/TPOT), \textit{tuning-only} improves FP16 by $1.33$--$1.55\times$. In multiple cases SLOs were unmet without optimization at lower TP, but became feasible after tuning and/or quantization (Tables \ref{tab:results}, \ref{tab:all-per-tp}). 
When workloads operate near stability boundaries, the Benchmarker+Tuner (exponential search + TPE) finds SLO-compliant regions with higher sustainable rates, so relative tuning gains are largest in the most latency-critical production cases. In throughput-focused regimes, we observed diminishing returns, which warrant further investigation.

\paragraph{ROI of automation: Amortizing Siloed Efforts} We quantify in Figure \ref{fig:page1-figure} the engineering cost of manual optimization, estimating it at 80--100 hours of specialized effort, compared to 15--25 hours for an automated \textsc{OptiKIT} run. For an industry setting, the primary contribution is not just the throughput gain but the drastic reduction in specialized, manual engineering cost. Manual optimization is a source of \textit{hidden complexity} that creates knowledge silos, non-reproducible artifacts, and duplicated efforts across teams. By standardizing the process into a reproducible, end-to-end pipeline, \textsc{OptiKIT} democratizes performance tuning, enabling any application team to achieve expert-level optimization and directly addressing the organizational bottleneck of scarce ML systems expertise.

\begin{figure}[t]
    \centering
    \includegraphics[width=\linewidth]{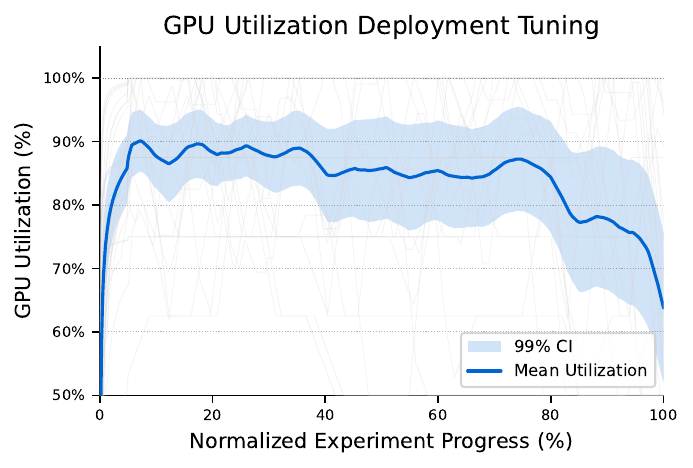}
    \caption{\textbf{GPU Utilization during Deployment Tuning.} Mean utilization across concurrent trials (88.0\%); shaded region shows 99\% CI. The drop near end highlights ``last-wave'' inefficiency from synchronization barriers.}
    \label{fig:gpu_utilization}
\end{figure}

We also quantify computational efficiency: Figure~\ref{fig:gpu_utilization} shows a mean GPU utilization of 88.0\% during deployment tuning, with the 12\% idle fraction stemming from (i) \textit{within-wave straggler variance}, where some trials take longer to reach steady-state, leaving other GPUs idle at the barrier, and (ii) \textit{last-wave under-utilization} when remaining trials are not divisible by available GPUs.

\paragraph{Risk Mitigation \& Failure Modes}
\textsc{OptiKIT} addresses failure modes at multiple levels.
First, the \textit{StatEval} quality-gate (Figure \ref{fig:main_fig}) terminates pipelines whose quantized models fail accuracy checks before any expensive benchmarking or tuning begins; future versions will support automatic re-compression with varied calibration data or precision settings.
Second, cost blowups under tight SLOs are controlled via (a) early infeasibility detection that quickly prunes unviable configurations, (b) capped exponential-bracketing and bisection within a fixed trial budget to efficiently locate the feasible performance boundary, and (c) an SLO-compliant objective normalized per-GPU to maximize throughput while strictly satisfying latency constraints.
Third, tuning instability is handled on two fronts: the Benchmarker monitors steady-state regression signals (Eq.~\ref{eq: 1}), immediately classifying configurations with growing backlogs as SLO-failing, while strict actor-pool isolation between stages prevents operational conflicts and guarantees a clean GPU state for each trial.

Since the pipeline archives all trial metrics, extracting a multi-objective Pareto frontier over cost, latency, and quality is a straightforward post-processing step. Teams can analyze trade-offs to select alternative configurations if business requirements change, avoiding the need to re-run expensive optimization sweeps.

\paragraph{Scope Limitations} Our current evaluation covers dense decoder-only models with standard decoding and tensor parallelism up to 8 GPUs. We do not yet evaluate MoE-specific decision variables, speculative decoding parameters, or stateful multi-turn session dynamics (e.g., persistent KV-cache reuse and cross-turn dependencies).

\section{Conclusion \& Future Work}

\subsection{Conclusion}

We present \textsc{OptiKIT}, an end-to-end, production-grade framework for automated LLM optimization with distributed and dynamic resource management. Unlike existing toolchains that address isolated aspects of the process, \textsc{OptiKIT} automates every stage---from model fetching and compression to statistical evaluation, inference benchmarking, and deployment tuning---enabling enterprise teams to obtain a production-ready optimized model with minimal manual intervention.
Empirical evaluations across diverse model families and real-life production configurations demonstrate more than $2\times$ throughput improvements per GPU while maintaining near full-precision accuracy across reasoning, instruction-following, and safety benchmarks. These results validate \textsc{OptiKIT} as an effective and reproducible solution for large-scale, production-grade LLM optimization.

\subsection{Future Work}
Key directions for future development include:
\textit{(i) Advanced compression:} Integrating pruning~\cite{lecun1989optimal, han2015deep, zhang2023dynamic, WandaSunetal2024}, structured sparsity~\cite{ma2023llm, mishra2021acceleratingsparsedeepneural}, and sparse-quantized representations~\cite{dettmers2023spqr} alongside quantization, which will require co-optimization strategies beyond the current sequential pipeline.
\textit{(ii) Asynchronous scheduling:} Replacing the current inter-stage synchronization barriers with an async task scheduler to improve GPU utilization and eliminate last-wave idle time (Section~\ref{sec:final_algorithm}).
\textit{(iii) MoE and speculative decoding:} Extending to Mixture-of-Experts (expert parallelism, EP/TP co-tuning) and speculative decoding (draft model selection, acceptance policy), both requiring dedicated quality-latency-cost studies.
\textit{(iv) Multi-turn workloads:} Adding trace-driven, stateful benchmarking with cache-aware objectives that account for KV-cache residency, eviction pressure, and session-level tail latency.

\section{Disclaimer}
This research was conducted at eBay Inc. All intellectual property arising from this work is the sole property of eBay Inc. The external collaborator’s involvement was limited to academic discussion and manuscript preparation and does not confer any ownership or intellectual property rights. All software, data, and methodologies described in this work were created under eBay’s direction and within its research environment.

%% file: sections/0X_appendix.tex
\section{System Design and Architecture} \label{app: System Design and Architecture}

\subsection{Architecture Components} 

This appendix provides an illustrative view of the core OptiKIT runtime components. Each snippet corresponds to a minimal, self-contained example that demonstrates how optimization flows are constructed, submitted, and executed within the distributed optimization framework.

\subsection{Code walkthrough and explanations}

\paragraph{Optimizer \& Recipe definition.}

The first example shows the high-level optimizer interface. The user selects a backend implementation (here the vLLM compressor) and instantiates a quantization recipe. Recipes encapsulate all quantization hyperparameters and produce a concrete execution strategy through the create() call. The pipeline then runs end-to-end—from model retrieval to compression and artifact generation—under a unified interface once a backend adapter is implemented.

\begin{lstlisting}[language=Python]
# Unified interface across implemented backends
optimizer = Optimizer(LLMCompressorBackend())
# vLLM backend (used in this paper)

# Recipe bundles all optimization specifications
recipe = get_recipe("int_w8a8")  
# W8A8 quantization recipe
strategy = recipe.create()        
# Complete configuration generated

# Automated optimization pipeline
optimizer.run_pipeline(
    model_path="llama-70b/v1.0",
    output_path="./optimized_model", 
    strategy=strategy
)
\end{lstlisting}

\paragraph{Flow definition.} 
The next listing defines a registered flow responsible for orchestrating quantization trials. A flow coordinates distributed actors in Ray, creates resource-scaled pools for quantization and evaluation, and manages trial queues as resources become available. Each flow explicitly declares its required parameters, enabling validation and reproducibility at submission time.

\begin{lstlisting}[language=Python]
@FlowRegistry.register("quantization")
class QuantizationFlow(BaseFlow):

    def run(self, job: OptimizationJob) -> Dict[str, Any]:
        
    # Create ActorPools with dynamic resource allocation
        quant_pool = self._create_quantization_actors(
            context
        )
        eval_pool = self._create_evaluation_actors(
            context
        )

    # ActorPools queue trials and consume them as actors become available
        self._run_quantization_stage(
            trials, quant_pool
        )
        self._run_evaluation_stage(
            trials, eval_pool
        )
        results = self._build_results(
            trials
        )
        return results

    @property
    def required_params(self) -> List[str]:
        return [
            "quantization_recipe",
            "num_trials"
        ]
\end{lstlisting}

\paragraph{Quantization Actor}

Each quantization actor performs one independent compression trial. It loads the model, applies the specified quantization recipe, and emits the path of the resulting optimized model. Actors are GPU-bound and execute in isolation, ensuring deterministic per-trial behavior and clean teardown between experiments.

\begin{lstlisting}[language=Python, label={lst:quantization-actor}]
@ray.remote
class QuantizationActor(BaseActor):

    def run(self, trial_id: str, config: QuantizationConfig):
        
        result = self._compress(
            config.model_path,
            config.quantization_recipe
        )

        return {"quantized_model_path": result}
\end{lstlisting}

\paragraph{Submission example.}

This submission example shows how an optimization job is described and dispatched. A job specification includes model metadata (from MMS), calibration dataset location, flow parameters such as the quantization recipe and number of trials, and hardware requirements. The submitter component serializes the configuration and triggers execution on the Ray cluster, returning structured results with metrics and artifact locations.

\begin{lstlisting}[language=Python, label={lst:submission-engine}]
job = OptimizationJob(
    name="llama_70b-compression-job",
    flow="quantization",
    model=MMSModelConfig(
        repo="models",
        name="llama-70b",
        version="v1.0"
    ),
    dataset=HadoopDatasetConfig(
         hdfs_path="/data/calibration"
    ),
    flow_params={
        "quantization_recipe": "int_W8A8 (Dynamic)",
        "num_trials": 5
    },
    compute_config=[
        ComputeConfig(sku=ResourceSKU.H100_8)
    ]
)

submitter = Submitter()
result = submitter.submit(job)
\end{lstlisting}

\newpage
\newpage

\section{Experimental results} 

All throughput measurements were collected using a steady-state inference benchmark based on the vLLM serving stack. Each configuration was tested under fixed input/output sequence lengths and latency SLOs as shown in the table headers. Reported values correspond to the normalized per-GPU TPS achieved while meeting the latency target. Runs lasted 900\,s of requests submission per sweep to ensure stable utilization. Configurations that did not satisfy latency SLOs are marked as $^{**}$.

\begin{table*}[!t] %
\centering
\caption{\textbf{Normalized per-GPU throughput for \textbf{FP16 tuning} across tensor parallelism levels.} Values are normalized per-GPU RPS (SLO-compliant). Gains are shown vs.\ FP16 baseline. Missing baselines ($^{**}$) indicate configurations not measured or not SLO-compliant.}
\vspace{-8pt}
\label{tab:fp16-tuning}
\renewcommand{\arraystretch}{1.15}
\begin{tabularx}{1\textwidth}{@{} 
    >{\centering\arraybackslash}X 
    >{\centering\arraybackslash}X 
    >{\centering\arraybackslash}X 
    >{\centering\arraybackslash}X @{}}
\toprule
\textbf{TP} 
  & \textbf{Baseline (FP16)} 
  & \textbf{FP16 (Tuned)} 
  & \textbf{Gain (Tuned / Baseline)} \\
\midrule
\multicolumn{4}{c}{\textbf{Qwen 2.5 7B (Input 1200, Output 80, Latency P95 500\,ms)}} \\
\midrule
1 & — & — & $^{**}$ \\
2 & 3.52 & 5.12 & 1.45$\times$ \\
4 & 4.68 & 6.79 & 1.45$\times$ \\
\midrule
\multicolumn{4}{c}{\textbf{Mistral Small 3 24B (Input 3000, Output 200, Prefix 2000, Latency P95 1500\,ms)}} \\
\midrule
1 & — & — & $^{**}$ \\
2 & — & — & $^{**}$ \\
4 & 0.604 & 0.937 & 1.55$\times$ \\
\midrule
\multicolumn{4}{c}{\textbf{Mistral Small 3 24B (Input 1500, Output 1500, Prefix 1000, TTFT P50 50\,ms; TPOT P50 10\,ms)}} \\
\midrule
1 & — & — & $^{**}$ \\
2 & — & — & $^{**}$ \\
4 & 0.562 & 0.750 & 1.33$\times$ \\
\bottomrule
\end{tabularx}
\vspace{2pt}
\raggedright
\footnotesize $^{**}$SLOs not met or FP16 baseline unavailable for the given TP.
\vskip -0.1in
\end{table*}

\begin{table*}[!t]
\centering
\caption{\textbf{Normalized per-GPU throughput and improvement vs.\ FP16 baseline across models, tensor parallelism, and bitwidths.} Values are normalized per-GPU RPS (SLO-compliant). Missing baselines ($^{**}$) indicate configurations not measured or not SLO-compliant.}
\vspace{-8pt}
\label{tab:all-per-tp}
\renewcommand{\arraystretch}{1.15}
\begin{tabularx}{\textwidth}{@{} c c 
    >{\centering\arraybackslash}X 
    >{\centering\arraybackslash}X 
    >{\centering\arraybackslash}X 
    >{\centering\arraybackslash}X 
    >{\centering\arraybackslash}X @{}}
\toprule
\textbf{TP} & \textbf{Bitwidth} 
  & \textbf{Baseline (FP16)} 
  & \multicolumn{2}{c}{\textbf{Quantization only}} 
  & \multicolumn{2}{c}{\textbf{Quantization + Tuning}} \\
\cmidrule(lr){4-5}\cmidrule(lr){6-7}
  & 
  & 
  & \textbf{Norm. TPS} & \textbf{Gain} 
  & \textbf{Norm. TPS} & \textbf{Gain} \\
\midrule
\multicolumn{7}{c}{\textbf{Qwen 2.5 7B (Input 1200, Output 80, Latency P95 500\,ms)}} \\
\midrule
1 & FP W8A8 (Dynamic)  & —    & 3.70 & $^{**}$ & 3.67 & $^{**}$ \\
1 & INT W8A8 (Dynamic) & —    & 3.66 & $^{**}$ & 3.67 & $^{**}$ \\
1 & INT W4A16 & —    & 2.01 & $^{**}$ & 3.20 & $^{**}$ \\
\midrule
2 & FP W8A8 (Dynamic)  & 3.52 & 5.96 & 1.69$\times$ & 7.49 & 2.13$\times$ \\
2 & INT W8A8 (Dynamic) & 3.52 & 5.95 & 1.69$\times$ & 7.50 & 2.13$\times$ \\
2 & INT W4A16 & 3.52 & 5.15 & 1.46$\times$ & 5.14 & 1.46$\times$ \\
\midrule
4 & FP W8A8 (Dynamic)  & 4.68 & 5.62 & 1.20$\times$ & 5.60 & 1.20$\times$ \\
4 & INT W8A8 (Dynamic) & 4.68 & 5.62 & 1.20$\times$ & 5.62 & 1.20$\times$ \\
4 & INT W4A16 & 4.68 & 4.68 & 1.00$\times$ & 4.68 & 1.00$\times$ \\
\midrule
\multicolumn{7}{c}{\textbf{Mistral Small 3 24B (Input 3000, Output 200, Prefix 2000, Latency P95 1500\,ms)}} \\
\midrule
1 & FP W8A8 (Dynamic)  & — & — & — & — & — \\
1 & INT W8A8 (Dynamic) & — & — & — & — & — \\
1 & INT W4A16 & — & — & — & — & — \\
\midrule
2 & FP W8A8 (Dynamic)  & —     & —     & $^{**}$ & 0.614 & $^{**}$ \\
2 & INT W8A8 (Dynamic) & —     & —     & $^{**}$ & —     & $^{**}$ \\
2 & INT W4A16 & —     & 0.335 & $^{**}$ & 0.617 & $^{**}$ \\
\midrule
4 & FP W8A8 (Dynamic)  & 0.604 & 1.732 & 2.87$\times$ & 1.734 & 2.87$\times$ \\
4 & INT W8A8 (Dynamic) & 0.604 & 1.523 & 2.52$\times$ & 1.523 & 2.52$\times$ \\
4 & INT W4A16 & 0.604 & 1.125 & 1.86$\times$ & 1.523 & 2.52$\times$ \\
\midrule
\multicolumn{7}{c}{\textbf{Mistral Small 3 24B (Input 1500, Output 1500, Prefix 1000, TTFT P50 50\,ms and TPOT P50 10\,ms)}} \\
\midrule
1 & FP W8A8 (Dynamic)  & — & — & — & — & — \\
1 & INT W8A8 (Dynamic) & — & — & — & — & — \\
1 & INT W4A16 & — & — & — & — & — \\
\midrule
2 & FP W8A8 (Dynamic)  & —     & —     & —      & 0.506 & $^{**}$ \\
2 & INT W8A8 (Dynamic) & —     & 0.221 & $^{**}$ & 0.506 & $^{**}$ \\
2 & INT W4A16 & —     & 0.148 & $^{**}$ & 0.492 & $^{**}$ \\
\midrule
4 & FP W8A8 (Dynamic)  & 0.562 & —     & —      & 0.875 & 1.56$\times$ \\
4 & INT W8A8 (Dynamic) & 0.562 & 0.531 & 0.95$\times$ & 0.875 & 1.56$\times$ \\
4 & INT W4A16 & 0.562 & 0.562 & 1.00$\times$ & 0.875 & 1.56$\times$ \\
\bottomrule
\end{tabularx}
\vspace{2pt}
\raggedright
\footnotesize $^{**}$SLOs not met or FP16 baseline unavailable for given TP.
\vskip -0.1in
\end{table*}